\documentstyle[prl,aps]{revtex}
\begin{document}

\input epsf
\draft
\twocolumn[\hsize\textwidth\columnwidth\hsize\csname
@twocolumnfalse\endcsname
\title{Vortex lattices in strong type-II superconducting two-dimensional strips}
\author{J. J. Palacios$^*$}
\address{Department of Physics and Astronomy, University of Kentucky, 
Lexington, KY 40506, USA.}

\date{\today}
\maketitle

\widetext
\begin{abstract}
\leftskip 2cm
\rightskip 2cm

We show how to calculate {\em semi-analytically} the dense vortex state in
strong type-II superconducting nanostructures. For the specific  case of a
strip, we find vortex lattice solutions which also incorporate
surface superconductivity. We calculate the energy cost to displace
individual vortex rows parallel to the surfaces and find that this energy
oscillates with the magnetic field. Remarkably, we also find that, at a
critical field $H^*$ below $H_{c2}$, this ``shear'' energy becomes strictly
zero for the surface rows due to an unexpected mismatch with the bulk
lattice.

\end{abstract}

\vskip2pc]

\narrowtext
Despite increased technological interest in the transport properties of the
vortex lattice state (VLS)\cite{Abrikosov} in type-II superconducting
nanostructures, the theoretical understanding of the interplay between the VLS 
and the interfaces is still mostly qualitative.  While experiments are
routinely done in nanostructures with various interface
geometries\cite{filmsexp,Pruymboom,Bezryadin} and explore all values of
magnetic field $H$ and temperature $T$, the order parameter, $\Psi({\bf r})$,
has been almost exclusively calculated in only two regimes. (i)  For fields
above the upper critical field $H_{c2}$, which separates the VLS from the
metallic phase, solutions of the Ginzburg-Landau differential equations
revealed long ago the existence of a highly localized order parameter at the
surfaces of the sample which can survive up to a higher critical field
$H_{c3}=1.695 H_{c2}$\cite{surfaceold}.  Recently, experiments in novel
engineered nanostructures\cite{Bezryadin} have spurred the interest in this
surface superconductivity and new calculations of upper critical fields have
been done for different systems\cite{Bezryadin,Buzdin}. (ii)  On the other
extreme, $\Psi({\bf r})$ has been calculated for values of $H$ just above the
lower critical field $H_{c1}$, which separates the Meissner state from the VLS.
The order parameter takes the form of a dilute VLS in which the inter-vortex
distance, $a_0$, is large compared to the coherence length $\xi$ (typical size
of the vortex cores).  The properties of this dilute VLS in the presence of a
surface parallel to $H$ were analyzed in the past\cite{Ternovskii} and numerous
studies in thin films have been reported over the years\cite{filmsexp}.

The dilute VLS, which is correctly described in the London approximation where
the vortex cores are ignored, only exists in a narrow range of $H$ above
$H_{c1}$ for strong type-II ($\kappa \gg 1$)
superconductors\cite{Tesanovic,Brandt}. Instead, a dense vortex state occupies
most of the $H-T$ phase diagram down to $H \approx 0.3 H_{c2}$. In this regime,
the vortex cores fill most of the space, interact strongly among themselves,
and, most importantly, interact directly with the interfaces. To the best of
our knowledge, only numerical simulations for  nanostructures such as thin
films or slabs (placed parallel to $H$)\cite{Bolech} and strips  (perpendicular
to $H$)\cite{Vinokur} have been reported in this regime. Our goal is to show
that, for systems with simple geometries and simple orientations with respect
to $H$, one can minimize {\em semi-analytically} the Ginzburg-Landau energy
functional and obtain the full solution of the order parameter in the dense
vortex regime.   We illustrate the procedure for a disorder-free
two-dimensional strip perpendicular to $H$. The order parameter comprises,
generically, both a dense VLS and enhanced superconductivity at the surfaces.
In addition,  this procedure allows us to calculate the energy cost to displace
individual rows parallel
 to the surfaces. This ``shear'' energy, $E_s$,  which is
directly related to the measurable flow stress\cite{Pruymboom}, is, however,
elusive in  numerical simulations\cite{Vinokur}.  

At the mean-field level (excluding thermal 
fluctuations\cite{Tesanovic}) the order parameter can be obtained from 
minimization of the Ginzburg-Landau functional
\begin{eqnarray}
G&=&G_n+\int d{\bf r} \left[ \alpha |\Psi({\bf r})|^2 + 
\frac{\beta}{2}|\Psi({\bf r})|^4 + \right.\nonumber \\
&&\left.\frac{1}{2m^*}\Psi^*({\bf r})\left(-i\hbar{\bf\nabla} - 
\frac{e^*}{c}{\bf A({\bf r})}\right)^2\Psi({\bf r}) +
\frac{[h({\bf r})-H]^2}{8\pi} \right],
\label{G-L}
\end{eqnarray}
where $G$ and $G_n$ are the Gibbs free energies of the superconducting and
metallic states, respectively. The term $[-i
\hbar \nabla - e^* {\bf A}({\bf r})/c]^2/2m^*$ is the kinetic energy
operator for Cooper pairs of charge $e^*=2e$ and mass $m^*=2m$ in a
vector potential ${\bf A}({\bf r})$ associated with the magnetic
induction $h({\bf r})$.  The parameters $\alpha$ and $\beta$ have the
usual meaning\cite{Tinkham}. 

\begin{figure}
\centerline{ \epsfxsize=7cm \epsfbox{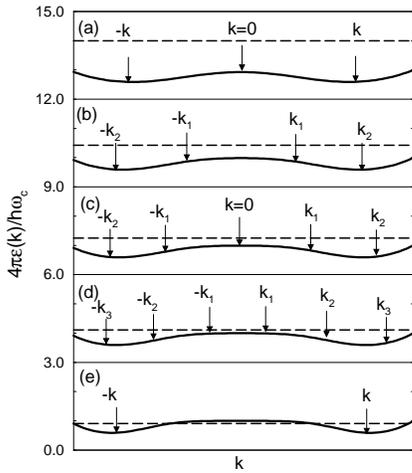}}
\caption{Cooper pair LB structure (in units of the cyclotron energy
$\hbar\omega_c/2$) in a $W=16\xi$ strip for different values of $H$ (shifted
for clarity). Notice the bending of the LB near the  surfaces. Dashed lines
represent $-\alpha$.  As $H$ increases [from (a) to (e)] the number of
components of the order parameter increases progressively.  When $-\alpha$ lies
below the LB in the center of the strip [panel (e)] only surface
superconductivity survives.}
\label{fig1}
\end{figure}

If, for simplicity, we restrict ourselves to two dimensions, $\kappa$ becomes
effectively infinite and $h({\bf r})=H$\cite{note}.  In order to be able to
find analytically the solutions which minimize Eq.\ \ref{G-L} we first expand 
the order parameter as $\Psi({\bf r})=\sum_{p,n} C_{p,n} \Phi_{p,n}({\bf r})$,
where $\Phi_{p,n}({\bf r})$ are the normalized eigenfunctions of the kinetic
energy operator.  The geometry of the interfaces determines the appropriate
gauge choice and quantum number $p$ (which is usually the linear or angular
momentum); the microscopic details of the interfaces determine the boundary
conditions for the eigenfunctions\cite{surfaceold,Bezryadin,Tinkham}.  To
obtain the order parameter in the dense vortex regime it is sufficient to
consider an expansion of $\Psi({\bf r})$ in the lowest band (LB),
$n=0$\cite{anothernote}. The LB expansion allows us to write
Eq.  (\ref{G-L}) as
\begin{eqnarray}
G-G_n&=&\sum_p^{N_c} \alpha_p |C_p|^2 + \nonumber \\
&& \frac{\beta}{2}
\sum_{p_1,p_2,p_3,p_4}^{N_c} C^*_{p_1}C^*_{p_2}C_{p_3}C_{p_4}
\int d{\bf r}\: \Phi^*_{p_1}\Phi^*_{p_2}\Phi_{p_3}\Phi_{p_4}
\label{LLL}
\end{eqnarray}
where $N_c$ is the number of components, $\alpha_p=\alpha+\epsilon(p)$ is the
condensation energy of the $p$ component, and the second  term represents the
``interaction'' between Cooper pairs. The interfaces are responsible for the
non-uniformity of the condensation energy through the kinetic energy,
$\epsilon(p)$, of the Cooper pair.   $-\alpha_p$ increases near interfaces with
insulators, which favors surface superconductivity (see Fig.\ \ref{fig1}),
and shrinks near those with metals . 

It is the $p$ dependence of $\alpha_p$ which, in principle, makes the
minimization of Eq.\ (\ref{LLL}) non-trivial.   In order to accomplish this
task one has to find: (i) the optimum $N_c$, (ii) the quantum numbers $\{p\}$,
and (iii) the complex coefficients $C_p$.  Some simple
considerations (which are specific for each geometry) allow us to solve this
problem analytically.   As an illustration we calculate the order parameter in
an isolated strip. The dimensions of this ideal strip are 
$L_y\rightarrow \infty$, $L_z \rightarrow 0$, and $W$, 
while $H$ is chosen in the
$z$ direction. The appropriate LB eigenfunctions are (in the Landau
gauge) $\Phi_k({\bf r}) \propto e^{iky}\chi_k(x)$, where $k$ is the wavevector
in the $y$ direction and $\chi_k(x)$ are nodeless functions which may be
calculated numerically subjected to the boundary conditions of zero current on
both surfaces. The integrals in Eq.\ \ref{LLL} become
\begin{eqnarray}
I(k_1,k_2,k_3,k_4)&\equiv& (L_y L_z)^{-1} \delta_{k_1+k_2,k_3+k_4}
\nonumber \\
&& \int dx \chi_{k_1}(x)\chi_{k_2}(x)\chi_{k_3}(x)\chi_{k_4}(x)
\end{eqnarray}
and may also be calculated numerically.

To find the lowest energy configuration we minimize  the Gibbs free energy in
Eq.\ (\ref{LLL}) for each possible $N_c$ separately (we will skip the trivial
case $N_c=1$, which is only relevant for very narrow strips with  $W\approx
\xi$\cite{Tinkham}):

\underline{One vortex row ($N_c=2$):} From the symmetry of the system one can
consider the two components in a pair $(k,-k)$ with $k > 0$ [see, e.g.,
Fig.\ \ref{fig1}(e)]. The symmetry also tells us that we can choose
$C_k=C_{-k}$. (Since the overall phase is irrelevant and the relative one only
determines the position of the row in the $y$ direction,  we consider both
coefficients to be real and positive.) There are only two types of  interaction
terms: (i) $I(k,k,k,k)=I(-k,-k,-k,-k)$, which represent the interaction between
Cooper pairs occupying the same state [in shorthand, $I(k)$],  and (ii)
$I(k,-k,-k,k)=I(k,-k,k,-k)=I(-k,k,-k,k)=I(-k,k,k,-k)$,  which represent the
interaction between Cooper pairs occupying different states [in
shorthand, $I(k,-k)$]. The Gibbs free energy becomes
\begin{eqnarray}
G-G_n&=&2 \alpha_k C_k^2 + \beta A_k C_k^4,
\end{eqnarray} 
where $A_k=I(k)+2I(k,-k)$.
Straightforwardly, one obtains the minimal values of the coefficients,
$\tilde C_k=\sqrt{-\alpha_k/\beta A_k}$, and the minimum Gibbs 
free energy, $\widetilde{G-G_n}=-\alpha_k^2/\beta A_k$,
for any pair $(k,-k)$.
Knowing the band structure and the interaction integrals,
the minimum-energy pair, $(\tilde k,-\tilde k)$, can now be found,
and $\Psi({\bf r})$ takes the form of one centered vortex row as long
as the two components overlap significantly\cite{twocomponents}.  For $W
\gg \xi$ and $H>H_{c2}$ the order parameter is {\em always} formed by
two components which do not overlap and constitute the usual surface
superconductivity on both sides of the strip [see Fig.\ \ref{fig1}(e)].

\underline{Two vortex rows ($N_c=3$):} In addition to the pair
$(k,-k)$ with positive, real coefficients $C_k=C_{-k}$
a third component can be considered at $k=0$ [see, e.g., Fig.\
\ref{fig1}(a)] with a complex coefficient $|C_0|e^{i\phi_0}$.
The Gibbs free energy becomes
\begin{eqnarray}
G-G_n&=&2 \alpha_k C_k^2 + \alpha_0|C_0|^2 + 
\frac{\beta}{2} \{2 A_k C_k^4 + I(0)|C_0|^4 + \nonumber \\
&& 4 [2 I(k,0) +\cos{(2\phi_0)} I(k,0,-k)]C_k^2 |C_0|^2 \},
\label{two}
\end{eqnarray}
where a new type of interaction terms, $I(k,-k,0,0)=I(-k,k,0,0)=I(0,0,k,-k)=
I(0,0,-k,k)$, appear [$I(k,0,-k)$ in
shorthand]. The last term in Eq.\ \ref{two} represents the
``correlation" energy between rows and,
regardless of the magnitude of the
coefficients, the minimum energy is obtained for
$\phi_0=\pi/2$, i.e., when the two rows 
are offset by half a row period, $a^r_0/2$.  Upon minimization  one obtains
\begin{eqnarray} 
\tilde C_k&=&\sqrt{\frac{\alpha_0 S_k-\alpha_k I(0)}
{\beta[I(0)A_k-2S_k^2]}} \\
|\tilde C_0|&=&\sqrt{\frac{2\alpha_k S_k -\alpha_0 A_k}
{\beta[I(0)A_k-2S_k^2]}} \\
\widetilde{G-G_n}&=&\frac{-2\alpha_k^2I(0)-\alpha_0^2A_k+4\alpha_k
\alpha_0 S_k} {2\beta[I(0)A_k-2S_k^2]}
\end{eqnarray} 
for any trio $(-k,0,k)$, where $S_k=2I(k,0)-I(k,0,-k)$. 

\underline{Three vortex rows ($N_c=4$):} 
The four components may correspond to pairs
$(k_2,-k_2)$ and $(k_1,-k_1)$ with $k_2 > k_1 > 0$ [see Fig.\
\ref{fig1}(b)]. We assume $C_{k_2}=C_{-k_2}$ and
$|C_{k_1}|=|C_{-k_1}|$. Once again two of the phases have been set to
zero ($\phi_{k_2}$ and $\phi_{-k_2}$). 
The other two, $\phi_{k_1}$ and $\phi_{-k_1}$, must be chosen to 
minimize the energy. To make things analytically accessible
we drop, for the moment, the weak correlation terms between vortex rows
in the resulting Gibbs free energy:
\begin{eqnarray}
G-G_n&=&2 \alpha_{k_1} |C_{k_1}|^2 + 2 \alpha_{k_2}
|C_{k_2}|^2 + \nonumber \\
&& \beta ( A_{k_1} |C_{k_1}|^4 +  A_{k_2} |C_{k_2}|^4  +
4 S_{k_1,k_2} |C_{k_1}|^2 |C_{k_2}|^2 ),
\end{eqnarray}
where  $S_{k_1,k_2}=I(k_1,k_2)+I(k_1,-k_2)$. Minimizing we obtain
\begin{eqnarray} 
|\tilde C_{k_1}|&=&\sqrt{\frac{\alpha_{k_1}A_{k_2}-
2\alpha_{k_2}S_{k_1,k_2}} {\beta[4S_{k_1,k_2}^2-A_{k_1}A_{k_2}]}} \\
\tilde C_{k_2}&=&\sqrt{\frac{\alpha_{k_2}A_{k_1}-
2\alpha_{k_1}S_{k_1,k_2}} {\beta[4S_{k_1,k_2}^2-A_{k_1}A_{k_2}]}} \\
\widetilde{G-G_n}&=&\frac{\alpha_{k_1}^2A_{k_2}+\alpha_{k_2}^2A_{k_1}
-4\alpha_{k_1}\alpha_{k_2}S_{k_1,k_2}}
{\beta[4S_{k_1,k_2}^2-A_{k_1}A_{k_2}]}
\label{4com}
\end{eqnarray}
for any set of four components $(-k_2,-k_1,k_1,k_2)$. The correlation
terms between adjacent rows, which ultimately 
determine the relative position between them, can be recast in the form 
\begin{eqnarray} 
&\delta_{k_2,3k_1} 2\beta|C_{k_1}|^3 C_{k_2} 
I(-k_1,k_1,k_2)& \times \nonumber \\
&[\cos{(2\phi_{k_1}-\phi_{-k_1})} + \cos{(2\phi_{-k_1}-\phi_{k_1})}]&.
\end{eqnarray}
Notice that for $k_2=3k_1$, the three rows
have the same period $a^r_0$. The minimum contribution of this term to the energy 
corresponds to $\phi_{k_1}=\phi_{-k_1}=\pi$,  which sets the offset between
adjacent rows to $a^r_0/2$. The total energy can now be approximately obtained
by subtracting the ``lock-in'' energy $4\beta|\tilde C_{k_1} |^3 \tilde
C_{k_2}I(-k_1,k_1,k_2)$ from Eq.\
(\ref{4com})\cite{newnote}. If $k_2\ne3k_1$ the choice of phases is
irrelevant since the period of adjacent rows is different and they cannot fall
into place.   The above choice of phases gives the worst relative position
between non-adjacent rows, i.e., aligns non-adjacent vortices.  However, the
term that accounts for such correlation is proportional to
$I(-k_2,-k_1,k_1,k_2)$ and can be safely dropped since the overlap of
$\chi_{-k2}$ and $\chi_{k2}$ is negligible in the expected minimal
solution.

\begin{figure}
\centerline {\epsfxsize=7cm \epsfbox{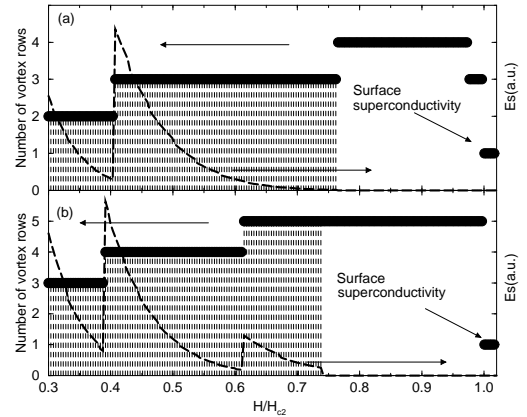}}
\caption{Number of vortex rows and $E_s$ as a function of $H$
for a strip with (a) $W=12\xi$ and (b)  $W=16\xi$. Shaded areas
correspond to values of $H$ with lock-in between the surface
rows and the bulk lattice. Non-shaded areas to values of $H$ without lock-in.}
\label{fig2}
\end{figure}

For more than three vortex rows one adds more components [see panels (c)
and (d) in Fig.\ \ref{fig1}] and follows similar considerations as for
$N_c=4$: (i) to neglect correlations between non-adjacent rows, (ii) to
minimize without the correlation terms for adjacent rows, and (iii) to
subtract the lock-in energy at the end.  One can thus obtain analytical
expressions for the $\tilde C_k$'s  and $\widetilde{G-G_n}$ for any
number of rows from which the optimum set of quantum
numbers $\{\tilde k\}$ can be found. We will only discuss the results in
the following.  Figure \ref{fig2} shows the number of vortex rows for
strip thicknesses $W=12\xi$ and $W=16\xi$ as a function of $H$.  As
expected, the number increases in a step-wise manner. Figure \ref{fig1}
shows several snapshots of $\epsilon(k)$ for the $W=16\xi$ strip for some
characteristic values of $H$.  As long as $-\alpha$ or ``chemical
potential" for the Cooper pairs (dashed lines) remains above the LB in
the center of the strip [panels (a) to (d)], the number of vortex rows
increases with $H$. When $-\alpha$ lies below the LB in the center,
i.e., when  $H>H_{c2}$ [panel (e)], the Cooper pairs can only nucleate
near the surfaces (surface superconductivity). For the narrower strip [Fig.
\ref{fig2}(a)] the number of vortex rows that can be accommodated before
the quenching of bulk superconductivity is logically smaller than for
the wider one [Fig.  \ref{fig2}(b)].  (To the numerical accuracy of our
calculations all the rows forming the lattice disappear at $H=H_{c2}$ in
the limit $W\rightarrow \infty$.) Peaks in the measured magnetization of
thin films have already been associated with these type of
transitions in the number of vortex rows\cite{filmsexp,Bolech}.

In contrast to the numerical simulations\cite{filmsexp,Bolech,Vinokur},
our analytical minimization allows us to calculate the energy cost
to ``shear'' or displace individual vortex 
rows in the direction of the surfaces. 
This energy $E_s$ is sample-dependent and proportional to the shear
modulus $c_{66}$ of the vortex lattice. To calculate $E_s$
one compares the minimum energy with that
obtained by switching the phases away from their optimum values so that
vortices in adjacent rows become aligned with each other.  Dashed lines in
Fig.\ \ref{fig2} correspond to $E_s$ for the surface 
rows.  Whenever a new row is added to the strip, $E_s$
experiences a sudden jump. This is because the addition of a new row
"squeezes" all the others against each other,
increasing their shear energy. 
On qualitative arguments\cite{Pruymboom,Moore}
 only one of the two possible lattice orientations is expected in
this system. Although near the transition points
the lattice is fairly distorted with  
$a^r_0\ne a_0$ ($a_0$ corresponds to the perfect triangular lattice), 
the expected orientation is still apparent in our results.
The experimental value of the flow stress,
which is proportional to $E_s$, has been reported in Refs.\
\onlinecite{Pruymboom} although instead of having the VLS confined in
strips they had it ``trapped" in disorder-free grooves etched in a
dirty film.  Oscillations in the flow stress were
reported for the whole range of $H$ and they were attributed to
dislocations that appear when the periodicities of the groove
superlattice and the VLS did not match. This picture does not consider
the possible overall distortion of the lattice. From our calculations we
conclude that, since the whole lattice reconstructs abruptly at each row
addition, dislocations need not be invoked to explain jumps in the flow
stress.  Note that the boundary conditions that mimic the  groove
interfaces are different from the ones used here, but, as long as the
vortex-interface interaction is sufficiently strong, this basic
conclusion holds.

\begin{figure}
\vspace{-3cm}
\hspace{-8.5cm}
\centerline{\epsfxsize=8cm \epsfysize=9cm \epsfbox{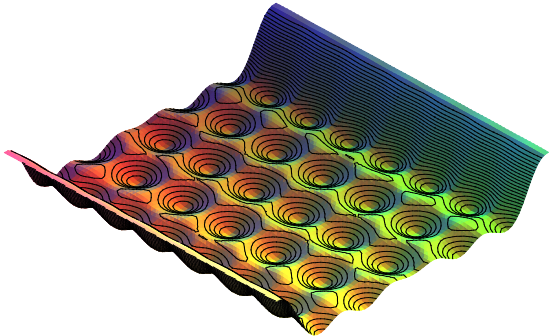}}
\vspace{-1cm}
\caption{Order parameter for a $W=16\xi$ strip
at $H=0.9H_{c2}$. Note the difference in period between the
surface vortex rows and the bulk lattice.}
\label{fig3}
\end{figure}

Finally, the most remarkable result to emerge from our calculations is
the fact that $E_s$ goes strictly to zero at a critical
field $H^*\approx 0.73H_{c2}$ as $W\rightarrow\infty$. This is due to the
fact that the surface rows become incommensurable with the bulk lattice 
at this critical field.  
Shaded areas in Fig.\ \ref{fig2} correspond to the situation
where lock-in takes place between the two surface vortex rows and the bulk
lattice and, consequently, there is always a finite
value of $E_s$ for such rows. For $H>H^*$ up to $H_{c2}$
the condensation energy near the surfaces always 
overcomes the lock-in energy.   The surface rows  prefer
to have a shorter period than the bulk ones
despite of the fact that lock-in is no longer possible.
Figure \ref{fig3}  shows the order parameter for $H$ near $H_{c2}$ where
the mismatch in period between surface rows and bulk lattice 
is apparent. This mismatch might reflect in
a premature quenching of the flow stress measured in deeply etched
NbGe grooves in experimental setups similar to those of Refs.\
\onlinecite{Pruymboom}.  The deep etching would create
the necessary spatial modulation of the condensation energy which
triggers this phenomenon.

The author thanks A. H. MacDonald for invaluable help in the early
stages of this work. The author also thanks H. Fertig, C. Brogan, and J.
Sinova for critical comments. This work has been funded by NFS Grants
Nos.  DMR-9416902 and DMR-9503814.


\begin{references}
\bibitem{anothernote} Away from the interfaces the LB approximation is
 equivalent to the lowest
Landau level approximation, but near the interfaces the LB
expansion includes higher Landau levels.
\bibitem{newnote} The relative error made in this approximation
 is typically less than 1\% as can be checked for the $N_c=3$ case, which is
 exactly solvable.
\end{references}
\end{document}